# Information-Theoretic Spectroscopy: Universal Sparsity of Extinction Manifold and Optimal Sensing across Scattering Regimes

Proity Nayeeb Akbar*

Department of Physics, Wesleyan University, 265 Church Street, Middletown, CT 06459-0155, USA

*Email: pakbar@wesleyan.edu

## Abstract

The inverse reconstruction of material properties from optical extinction efficiency ($Q_{ext}$) is constrained by the high-dimensional nature of Mie scattering spectra. We demonstrate that the $Q_{ext}$ manifold possesses an intrinsic, physics-governed sparsity that is universal across dielectric materials. By analyzing the spectral topology of a diverse polymer library, we identify a critical information bottleneck at the onset of the Mie transition ($r \approx 0.1\ \mu m$), where a peak in spectral entropy signifies a fundamental limit on signal compressibility. While the fast Fourier transform (FFT) is conventionally used for spectral analysis, we show it is physically mismatched for this domain; its periodic boundary assumptions induce spectral leakage that forces a massive expansion in the required basis set to resolve fine-scale Mie ripples. Conversely, the discrete cosine transform (DCT) mirrors the non-periodic geometry of extinction profiles, uncovering the inherent compressibility of $Q_{ext}$ by capturing over 90% of signal energy using fewer than 10 harmonic modes. Even at the Mie bottleneck, the DCT maintains a 12 × compression advantage over the FFT at a 99% energy threshold. Notably, while both bases converge to identical error floors for a fixed energy threshold, the DCT achieves this fidelity with significantly lower hardware overhead; at the information bottleneck, the top 10 DCT modes capture ∼37% of the signal energy, compared to only ∼19% for the FFT. Stress-testing under 10% additive Gaussian noise confirms that the characteristic information bottleneck is spatially and structurally invariant, proving that this complexity peak is a fundamental physical constant of the extinction manifold rather than a numerical artifact. By mapping this sparsity onto a compressed sensing architecture, we resolve a $2.5 - 20\ \mu m$ spectral range using between 22 sensors (Rayleigh/Geometric regimes) and 170 sensors (critical Mie transition)—enabling a $51\% - 94\%$ reduction in hardware complexity that breaks the traditional Nyquist sampling limit (350 sensors) for high-fidelity clinical and remote sensing applications.

## Introduction

Optical extinction efficiency ($Q_{ext}$) fundamentally encodes the complex interaction between light and particulate media, serving as a cornerstone observable in scattering and absorption spectroscopy [1, 2]. While traditionally analyzed through physical parameters—such as resonant modes, refractive indices, and size parameters [1, 2]—the underlying information-theoretic structure of $Q_{ext}$ remains largely unexplored. The quest to quantify this information dates back to Spänkuch [3], who qualitatively identified "sensitive regions" in extinction measurements where the information content regarding particle size is maximized. However, a rigorous quantitative explanation for why these regions exist remained elusive. This gap in understanding prompts a critical question for modern spectroscopy: What are the fundamental limits of information density within a $Q_{ext}$ spectrum, and how does this complexity scale across the diverse dielectric media?

We demonstrate that the answer to this scaling lies in a universal information bottleneck that occurs at the onset of the Mie transition ($r \approx 0.1\ \mu m$). At this peak of spectral entropy, the transition from Rayleigh power-law scaling to complex interference patterns reaches its maximum structural complexity [4]. This study defines the information bottleneck as the critical coordinate where the sparsity assumptions required for compressed sensing are most severely tested [5, 6, 7].

This bottleneck represents the maximum number of effective degrees of freedom required to represent a material's optical response [8, 9]. While traditional estimates of the number of degrees of freedom often overlook the impact of material contrast, recent analysis has shown that the degree of freedom is intrinsically linked to the scatterer's dielectric properties in nonlinear inverse problems [10]. Interestingly, our previous work demonstrated that model biological cells and 7-component polymer mixtures could be deconvolved from a single $Q_{ext}$ spectrum with near-perfect accuracy [11, 12, 13, 14]. While such high-fidelity reconstruction of high-dimensional mixtures appeared to surpass expected information-theoretic limits, by accounting for this contrast-dependent dimensionality, we reveal that the $Q_{ext}$ manifold is characterized by an inherent low-dimensionality. Consistent with the theory of structural invertibility [15], we show that the system's "effective dimension" matches the available degree of freedom at the Mie transition, providing the mathematical justification for the success of such counter-intuitive inverse reconstructions.

In this regime, the choice of a mathematical basis is not merely an optimization but a prerequisite for physical integrity [16]. A non-compressible representation results in an irrecoverable loss of information, exacerbating the ill-posed nature of inverse problems [13, 14, 11, 12] and leading to numerical instabilities that confound modeling [17, 18, 19]. Unlike data-driven models, such as Principal Component Analysis (PCA) [20] or Proper Orthogonal Decomposition (POD) [21], which require extensive training and lack a closed-form physical interpretation, or Wavelets [22] that excel at local discontinuities, $Q_{ext}$ profiles require a basis that can resolve global harmonic oscillations without introducing artifacts. We show that the conventional fast Fourier transform (FFT) [23] is physically mismatched for this task; its periodic boundary assumptions induce spectral leakage [24] that forces a massive expansion in the basis set to resolve fine-scale Mie ripples. Standard windowing techniques [25, 26] or parametric methods [27] can mitigate this leakage for frequency identification, but they remain suboptimal for signal reconstruction, as they either suppress the physically meaningful oscillations at the boundaries or require *a priori* assumptions about the underlying model. While experimentalists have successfully used the FFT-based compressed sensing to accelerate nano-FTIR imaging [5], we find that these sub-sampling strategies could be further optimized by adopting a basis that aligns with the non-periodic geometry of the extinction profiles.

Conversely, we establish that the discrete cosine transform (DCT) [28, 29] provides a physically consistent solution to this information bottleneck by mirroring the non-periodic geometry of extinction profiles. Our finding that the DCT basis more efficiently captures the Mie transition manifold aligns with recent advances in computer vision, where object detection is performed directly in the DCT domain to enhance speed [30]. By eliminating boundary-induced artifacts, the DCT uncovers the intrinsic sparsity of the $Q_{ext}$ manifold. We find that even at the Mie transition bottleneck, the DCT maintains a 12-fold efficiency advantage over the FFT, capturing 99% of signal energy with significantly fewer modes. Specifically, the top 10 DCT modes capture $\sim 37\%$ of the signal energy, nearly doubling the $\sim 19\%$ captured by the FFT. This provides a natural regularization for inverse problems: by concentrating signal energy into low-order DCT coefficients, we bypass the need to process the hundreds of spurious high-frequency terms introduced by Fourier leakage.

Critically, the utility of a spectral basis is defined not only by its sparsity in ideal conditions but by its stability under experimental stochasticity. We subject our framework to a rigorous sensitivity analysis, demonstrating that the DCT's energy compaction advantage is maintained even under 10% additive Gaussian noise. Importantly, we find that the characteristic information bottleneck at the Mie transition remains spatially and structurally invariant under noisy conditions, confirming it as a fundamental topological feature of the extinction manifold rather than a noise-sensitive artifact. This resilience confirms that the DCT coefficients align with the true physical signal of the Mie manifold, whereas the FFT's reliance on high-frequency leakage modes makes it fundamentally hypersensitive to noise. This robustness ensures that the transition from theoretical modeling to real-world hardware—where noise floors are a persistent constraint—does not compromise reconstruction fidelity.

Finally, we translate these insights into a hardware-software co-design principle. By mapping the DCT's sparsity onto a compressed sensing architecture derived from the condition number of the measurement matrix [6], we resolve a $2.5 - 20\ \mu m$ spectral range to capture 95% of the spectral energy using as few as 22 sensors (Rayleigh/Geometric regimes) to 170 sensors (critical Mie transition). This $51\% - 94\%$ reduction in hardware complexity compared to traditional Nyquist-sampled instruments ($\sim 350$ sensors) establishes the DCT as the optimal analysis framework to bypass the information bottleneck, enabling the next generation of smart sensors for real-time material characterization [31] in remote sensing [32] and clinical cytology [33].

# Methods

## Information-Theoretic Framework: Manifold Sparsity and Mie Bottleneck

The foundational premise of this methodology is that the $Q_{ext}$ spectrum is not a collection of arbitrary data points, but a structured manifold whose complexity is governed by the underlying physics of light-matter interaction. We hypothesize that this manifold resides in a low-dimensional subspace, the cardinality of which is limited by a universal information bottleneck at the onset of the Mie transition.

By identifying this bottleneck, we transform the one-time computational cost of high-resolution Mie simulations into a permanent reduction in experimental and computational complexity. The framework (schematized in **Figure 1**) operates through an integrated pipeline that maps physical scattering regimes to optimal sensing configurations:

1. A high-resolution Mie simulation is executed once to generate the $Q_{ext}$ spectrum ($N_{high}^{(\lambda)} \approx 620$ points). This establishes the spectral ground truth and allows for the calculation of the spectral entropy, identifying the peak structural complexity at the Mie transition ($r \approx 0.1\ \mu m$).
2. The DCT is applied to the high-resolution data. Unlike periodic transforms, the DCT's even-symmetric boundary conditions align with the non-periodic nature of $Q_{ext}$, concentrating the signal's energy into a minimal set of low-order modes ($M$). This step isolates the physically meaningful information from the "basis inflation" caused by spectral leakage.
3. The optimal number of dominant modes, $M$ (where $M \ll N_{high}^{(\lambda)}$), required to satisfy a predefined energy retention threshold (e.g., 99%) varies as a function of the scattering regime, reaching its maximum at the information bottleneck. These modes define the reduced-order subspace of the extinction manifold.
4. The $Q_{ext}$ spectrum is reconstructed using only the $M$ dominant DCT coefficients. This produces a continuous $N_{high}^{(\lambda)}$-point spectral profile from a sparse $M$-element vector, creating a computationally simpler forward model for inverse iterations.
5. Finally, the established sparsity of the DCT is leveraged to design a sub-Nyquist experimental setup. By analyzing the sensing matrix, we identify $P$ ($P \geq 2M$) discrete physical wavelengths that are maximally informative of the $M$ DCT modes. This allows future experiments to resolve the full spectral signature using only $P$ sparse measurements, fundamentally bypassing the need for high-resolution spectrometers.
6. In essence, the high-resolution simulation acts as a predictive lookup table, informing the construction of both a computationally efficient inverse model and an experimentally minimal measurement setup.

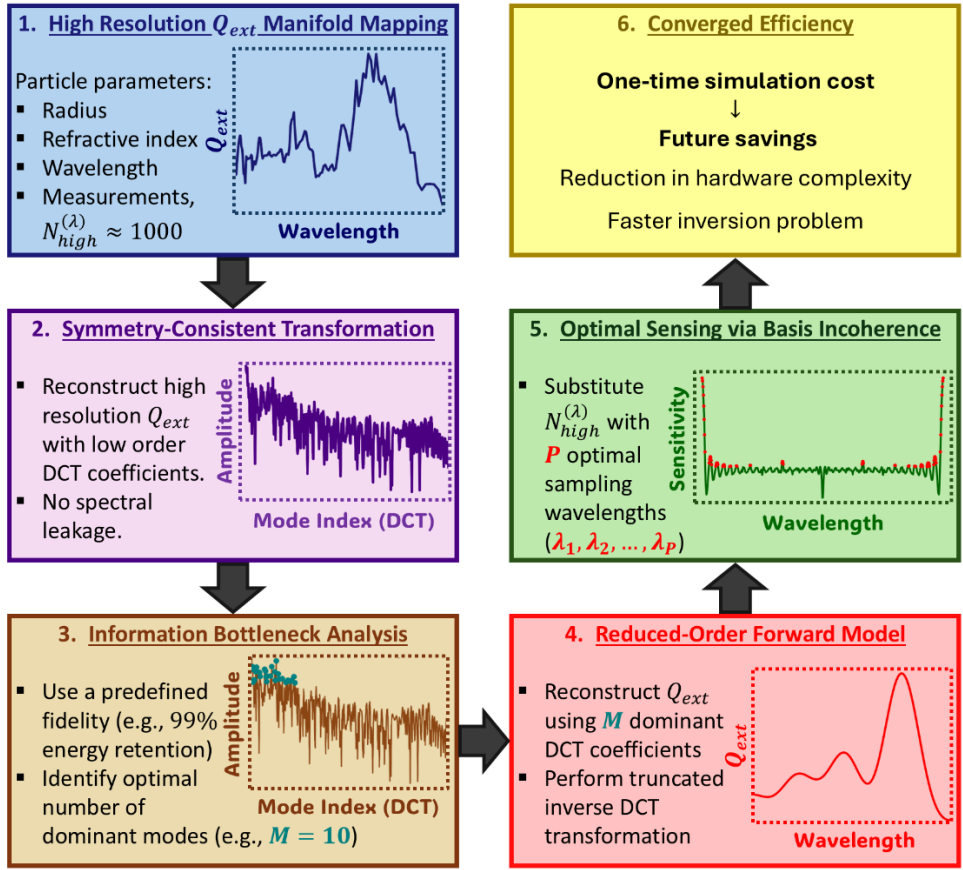

**Figure 1: Conceptual Framework for Information-Theoretic Spectroscopy.** The six-stage integrated workflow transforms high-resolution physical insights into experimental hardware optimization by exploiting the intrinsic sparsity of the extinction manifold. **(1) High-Resolution $Q_{ext}$ Manifold Mapping:** Initial characterization of the spectral ground truth ($N_{high}^{(\lambda)} \approx 620$) to establish the baseline signal complexity. **(2) Symmetry-Consistent Transformation:** Application of the DCT to map the non-periodic $Q_{ext}$ profile into a frequency domain that minimizes basis inflation and spectral leakage. **(3) Information Bottleneck Analysis:** Quantification of spectral entropy and dominant mode count ($M$) required to reach a target fidelity threshold (e.g., 99% energy retention), identifying the regime-dependent limits of signal compressibility. **(4) Reduced-Order Forward Model:** Inverse transformation of the $M$ dominant coefficients to generate a continuous spectral profile, providing a computationally lightweight engine for inverse modeling. **(5) Optimal Sensing via Basis Incoherence:** Identification of the $P$ discrete sampling wavelengths ($\lambda_1, \lambda_2, \ldots, \lambda_P$) that maximize information gain, enabling sub-Nyquist spectral reconstruction. **(6) Converged Efficiency:** The final framework yields a permanent reduction in hardware complexity and acquisition time, enabling real-time material characterization at the theoretical limits of information density.

## Forward Model Simulation and the $Q_{ext}$ Manifold Library

To investigate the universality of spectral sparsity, we constructed a comprehensive library of extinction efficiency ($Q_{ext}$) profiles for six homogeneous organic polymers: polymethyl methacrylate (PMMA), polycarbonate (PC), polydimethylsiloxane (PDMS), polyetherimide (PEI), polyethylene terephthalate (PET), and polystyrene (PS). These materials were strategically selected to represent a diverse range of complex dielectric functions, encompassing distinct refractive index dispersions and chemical backbones [34, 35, 36, 37, 38, 39]. This diversity ensures that the identified information-theoretic constraints are a result of fundamental scattering physics rather than material-specific absorption features.

The spectra were simulated for microspheres using Mie theory via the *PyMieScatt* package [40] across the mid-infrared band ($\lambda \approx 2.5 - 25\ \mu m$). To systematically map the evolution of the information bottleneck, we sampled a broad morphological parameter space with particle radii ranging from $r = 0.01\ \mu m$ to $25\ \mu m$. The spectral manifold is analyzed as a function of the dimensionless size parameter $x = 2\pi r/\lambda$. While the internal resonance structure is physically governed by the phase shift parameter $\rho = x\tilde{n}$ (where $\tilde{n}$ is the complex refractive index), using the material-independent $x$ allows us to demonstrate the universal nature of the information bottleneck across our polymer library.

This wide of range radii captures the transition through three distinct physical regimes: **1)** the Rayleigh regime ($x \ll 1$), where the spectra are governed by smooth, low-entropy power-law scaling; **2)** the Mie transition regime ($x \approx 1$), which is the locus of the critical information bottleneck, characterized by complex interference patterns and peak spectral entropy; and **3)** the Geometric regime ($x \gg 1$), where the signal converges toward a stable, high-frequency limit. By utilizing high-resolution complex refractive index data, the resulting $Q_{ext}$ manifold contains realistic features, including sharp molecular vibrations and broad scattering envelopes, providing a rigorous testbed for the stability and sparsity of the DCT basis.

## Spectral Signal Conditioning and Orthogonal Transformation

To ensure a standardized analysis of the extinction manifold, the simulated $Q_{ext}$ spectra were interpolated onto a uniform wavelength grid of $N = 620$ points. Before transformation, the signal mean was removed to isolate the oscillatory features of the Mie resonances from the DC offset, focusing the analysis on the structural complexity of the scattering profile.

We evaluated the compressibility of the extinction manifold using two distinct orthogonal bases: the discrete cosine transform (DCT) and the fast Fourier transform (FFT). This comparison is central to our hypothesis that the periodic assumptions of Fourier analysis are physically inconsistent with non-periodic optical spectra. To test if this inconsistency could be mitigated through standard signal processing techniques, the FFT was evaluated both in its raw form and in conjunction with a Hann window [41].

Critically, we hypothesized that windowing—while effective at suppressing spectral leakage in frequency identification—is inherently destructive for $Q_{ext}$ reconstruction. Because the boundary values of an extinction spectrum encode essential physical information regarding the scattering limits, the tapering effect of a Hann window attenuates meaningful oscillations, potentially degrading the integrity of inverse modeling. In contrast, the DCT's even-symmetric boundary conditions provide a naturally consistent basis that preserves boundary information without the need for artificial windowing.

The transformations were performed using the *SciPy* package [41], utilizing standard orthogonal normalization to ensure energy conservation between the spatial and transform domains:

$$\boldsymbol{F}^{(DCT)} = \boldsymbol{G}^{(DCT)}\{Q_{ext} - \bar{Q}_{ext}\}, \quad (1)$$
$$\boldsymbol{F}^{(FFT)} = \boldsymbol{G}^{(FFT)}\{Q_{ext} - \bar{Q}_{ext}\}, \quad (2)$$

where $\boldsymbol{G}^{(DCT)}$ and $\boldsymbol{G}^{(FFT)}$ represent the respective transformation operators and $\bar{Q}_{ext}$ is the signal mean. By projecting the extinction manifold into these harmonic subspaces, we can quantitatively define the information bottleneck through the lens of basis-dependent energy compaction.

## Quantifying the Information Bottleneck: Spectral Entropy

To identify the critical information bottleneck within the extinction manifold, we utilize spectral entropy ($\boldsymbol{H}$) as a quantitative measure of the structural complexity of the signal. In this context, $\boldsymbol{H}$ represents the statistical distribution of signal energy across the transformed modes; a low entropy value indicates that the signal's information is highly concentrated (sparse), while a high entropy value signifies that the energy is dispersed across the basis set due to either intrinsic physical complexity or basis-induced artifacts (spectral leakage).

The spectral entropy is calculated by applying the Shannon entropy definition [42] to the normalized power distribution ($\boldsymbol{P}_m^{(norm)}$) of the transform coefficients ($\boldsymbol{F}_m$):

$$\boldsymbol{P}_m^{(norm)}(\lambda) = \frac{|\boldsymbol{F}_m(\lambda)|^2}{\sum_{n=1}^{N}|\boldsymbol{F}_n(\lambda)|^2}, \quad (3)$$

$$\boldsymbol{H} = -\sum_{m=1}^{N} \boldsymbol{P}_m^{(norm)} \log_2\left(\boldsymbol{P}_m^{(norm)}\right), \quad (4)$$

where $\boldsymbol{P}_m^{(norm)}$, $\boldsymbol{F}_m$, and $\boldsymbol{H}$ represents the normalized power fraction, the transformation coefficient of the $m$-th mode, and the spectral entropy, respectively, for $m$ ranging from 1 up to $N$, the total number of modes ($N = 620$). Here, $\boldsymbol{F}_m$ represents either the DCT or FFT coefficients, i.e., $\boldsymbol{F}_m^{(DCT)}$ or $\boldsymbol{F}_m^{(FFT)}$, depending on the transformation being used.

By mapping $\boldsymbol{H}$ across the full range of particle radii ($0.01 - 25 \ \mu m$), we can pinpoint the Mie transition bottleneck where structural complexity reaches its global maximum. This metric allows us to contrast the "physical entropy" inherent in the scattering manifold against the "spurious entropy" introduced by the FFT's periodic boundary mismatch.

## Manifold Compression and Fidelity Scaling

To evaluate the compressibility of the extinction manifold and the efficiency of the harmonic bases, we employed two complementary truncation strategies. These methods were designed to characterize the minimum degrees of freedom required for physical representation and the rate of information decay under hardware constraints.

### Cumulative Energy Thresholding (The Sparsity Metric)

This approach identifies the intrinsic cardinality of the $Q_{ext}$ signal by determining the absolute minimum number of basis functions needed to represent the physics. For both the DCT and FFT, we calculated the normalized power of every coefficient, sorted them in descending order of magnitude, and summed them cumulatively.

We determined the smallest number of modes required ($m_{required}$) to reach strict fidelity thresholds of 90%, 95%, and 99% of the total spectral energy ($m_{90\%}$, $m_{95\%}$, and $m_{99\%}$). By treating $m_{required}$ as a regime-dependent variable, we can quantitatively map how the information bottleneck at the Mie transition forces an expansion of the required basis set. This metric establishes the theoretical bound for the degrees of freedom needed for a scientifically accurate inverse model.

### Sequential Mode Truncation (Fixed Mode Budget)

While thresholding establishes theoretical limits, sequential truncation evaluates the practical utility of a basis under a fixed computational or hardware budget. We assessed the energy compaction ($E_M$)—the percentage of total energy captured—when restricted to a predetermined number of dominant modes $M$ (specifically, $M = 5 - 100$).

This method answers a critical engineering question: Given a limited hardware budget, which basis maximizes information retention? Furthermore, this fixed-budget approach provides the basis for calculating various error metrics, allowing us to model how the reconstruction error decays as a function of the available sensing modes.

### Manifold Reconstruction

To validate the integrity of the compressed representations, reconstruction tests were performed by applying the inverse transformation ($G^{-1}$) to the truncated coefficient sets:

$$\hat{Q}_{ext} = G^{-1}\{F_{truncated}\} + \bar{Q}_{ext}, \tag{5}$$

The resulting signal, $\hat{Q}_{ext}$, represents the compressed extinction efficiency. By comparing $\hat{Q}_{ext}$ against the ground-truth Mie simulation across different scattering regimes, we can isolate the localized errors introduced by spectral leakage in the FFT versus the physically consistent reconstruction provided by the DCT.

## Quantitative Performance Metrics

To rigorously evaluate the fidelity and structural stability of the compressed $Q_{ext}$ manifold, we define metrics for energy compaction and reconstruction error. These metrics quantify how effectively each basis resolves the high-frequency Mie resonances that constitute the information bottleneck.

### Energy Compaction ($E_M$)

The total spectral energy ($E$) is defined as the sum of the power of all transformed coefficients. By the Parseval–Plancherel theorem [23], this energy is conserved between the physical domain ($Q_{ext}$) and the transform domain ($P_m$). We define energy compaction ($E_M$) as the percentage of total spectral energy retained within the first $M$ sequential modes:

$$E = \sum_{m=1}^{N} P_m, \tag{6}$$

$$E_M = \frac{\sum_{m=1}^{M} P_m}{E} \times 100\%, \tag{7}$$

This measures the ability of a basis to concentrate physical information into a low-dimensional subspace. A higher $E_M$ for a fixed $M$ indicates superior practical efficiency and a more physically consistent representation.

## Global and Local Fidelity Metrics

While energy retention is an indicator of sparsity, the quality of the inverse model depends on the distribution of reconstruction errors. We utilize three distinct metrics to assess the fidelity of the compressed signals ($\hat{Q}_{ext}^{(DCT)}$ and $\hat{Q}_{ext}^{(FFT)}$): **1)** root mean square error (RMSE), **2)** maximum absolute error ($E_{max}$), and **3)** local root mean square error (LRMSE).

### Root Mean Square Error (RMSE)

The root mean square error (RMSE) provides a global measure of average reconstruction fidelity across the entire $2.5 - 20\ \mu m$ range.

$$RMSE = \sqrt{\frac{1}{N}\sum_{i=1}^{N}(Q_{ext,i} - \hat{Q}_{ext,i})^2}, \tag{8}$$

### Maximum Absolute Error ($E_{max}$)

The maximum absolute error ($E_{max}$) identifies the worst-case deviation. High $E_{max}$ values typically signify non-physical artifacts, such as Gibbs phenomenon ringing near discontinuities [43], which occurs when a periodic basis (FFT) attempts to resolve the non-periodic boundaries of an extinction spectrum.

$$MAE = \max_{i}|Q_{ext,i} - \hat{Q}_{ext,i}|, \tag{9}$$

### Local Root Mean Square Error (LRMSE)

The local root mean square error (LRMSE) is a specialized metric designed to quantify reconstruction accuracy within the high-curvature Mie resonance peaks. Since these sharp features encode the particle's size parameter, their preservation is critical for robust inverse modeling. A peak-detection algorithm [41] is utilized with a relative prominence threshold that captures 2% of the maximum signal height to isolate significant physical resonances. For each identified peak, a localized spectral window is defined. The LRMSE is the average RMSE calculated exclusively within these high-complexity zones, revealing which basis better preserves the fine structure of the Mie manifold during the transition regime.

## Application: Optimal Hardware Co-Design via Basis Incoherence

The intrinsic sparsity of the $Q_{ext}$ manifold in the DCT domain provides the mathematical foundation for a transition from reduced-order modeling to experimental compressed sensing (CS). By identifying the minimal set of physical wavelengths ($P$) required to resolve the dominant $M$ modes, we can fundamentally reduce hardware complexity and acquisition time.

## Sensing Matrix and the Incoherence Principle

In the CS framework, the goal is to reconstruct the high-resolution spectrum $Q_{ext}$ from a sparse measurement vector $Y$, where $Y$ is obtained by sampling the full signal $Q_{ext}$ using a sparse measurement matrix $\Phi$:

$$Y = \Phi Q_{ext}, \qquad (10)$$

Given that $Q_{ext}$ is sparse in the DCT domain ($Q_{ext} = \Psi C$, where $\Psi$ is the inverse DCT operator and $C$ is the sparse coefficient vector), the reconstruction problem is formulated as:

$$Y = (\Phi\Psi)C = \Theta C, \qquad (11)$$

The matrix $\Theta = \Phi\Psi$ is the sensing matrix. Stable and robust reconstruction of the coefficient vector $C$ requires the measurement basis ($\Phi$) to be incoherent with the sparsity basis ($\Psi$). This ensures that the information contained in the dominant DCT modes is spread across the sampled physical wavelengths, satisfying the Restricted Isometry Property (RIP) [6].

### Sensitivity-Based Heuristic for Sensor Placement

To optimize sensor placement without the prohibitive computational cost of verifying the RIP condition for every possible configuration, we developed a sensitivity-based heuristic to identify the most informative sampling points:

1. Reduced subspace projection: We identify the $M$ indices of the dominant DCT modes that capture 99% of the spectral energy. We construct a reduced basis matrix, $\Psi_M$, by extracting the corresponding columns of the inverse DCT matrix.
2. Information sensitivity scoring: We define a sensitivity score ($S$) for each potential physical wavelength $\lambda_i$. This score quantifies the cumulative influence of the dominant modes on the measurement at that specific coordinate. A high $S$ indicates a wavelength that provides maximum leverage for resolving the signal's information-dense components:

$$S(\lambda_i) = \sum_{m=1}^{M} |\Psi_M(i,m)|, \qquad (12)$$

3. Optimal sensor selection: The $P$ physical wavelengths corresponding to the highest sensitivity scores are selected as the optimal, non-uniform sampling points for experimental acquisition.

By targeting these high-sensitivity regions, we ensure that the limited hardware budget (the $P$ sensors) is allocated to the specific spectral coordinates where the information bottleneck is most transparent, enabling high-fidelity reconstruction with a fraction of the traditional Nyquist-required data.

# Results and Discussion

## Spectral Characterization and Information Density

We first establish the numerical integrity of our framework. **Figure 2** (and **Figures S1-S6**) confirms that both DCT and FFT achieve perfect reconstruction of $Q_{ext}$ spectra for the six-polymer library when the full basis set is retained. This baseline confirms the mathematical completeness of both transforms, establishing that any subsequent variations in sparse recovery are a direct consequence of energy compaction efficiency rather than transform-induced error.

As the particle radius ($r$) increases beyond the Rayleigh limit ($r \geq 0.1\ \mu m$), the spectra transition from smooth, absorption-dominated envelopes to resonance-dominated profiles. The emergence of fine-scale ripple structures near $r \approx 0.1\ \mu m$ fundamentally shifts the physical dimensionality of the signal.

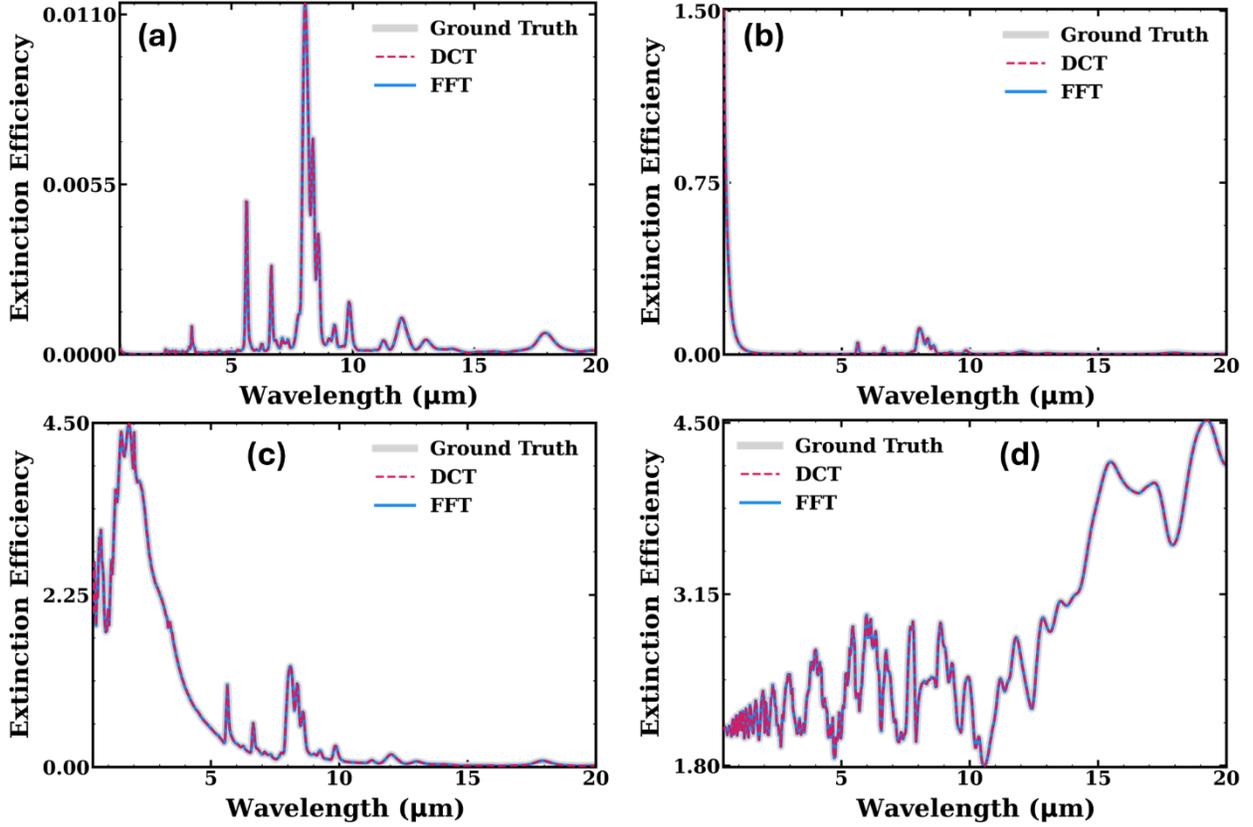

**Figure 2: Comparison of Reconstruction Fidelity Between DCT and FFT Bases.** Validation of the reconstructed spectra against the ground-truth interpolated $Q_{ext}$ for PC microspheres across varying size parameters: **(a)** $r = 0.01\ \mu m$, **(b)** $r = 0.1\ \mu m$, **(c)** $r = 1.0\ \mu m$, and **(d)** $r = 10.0\ \mu m$. In the absence of the coefficient truncation, both the DCT and FFT transformations maintain perfect numerical alignment with the ground-truth signal. This baseline confirms the mathematical completeness of both bases, establishing that any subsequent variations in sparse recovery performance (as seen in **Figure 4**) are a direct consequence of energy compaction efficiency rather than transform-induced error.

The spectral energy distributions in **Figures 3** and **S7-S12** illustrate this evolution across scattering regimes, with low-order modes ($M \leq 50$) displayed on a linear scale to highlight dominant features and the full spectrum on a semi-log scale to visualize the decay of high-frequency components. In the Rayleigh regime, energy is concentrated in a minimal set of leading coefficients, and the spectral response is characterized by low overall amplitude ($10^{-5}$ to $10^{-2}$). However, upon entering the Mie transition region, spectral amplitudes rise significantly ($10^{-3}$ to $10^{1}$), and the interference between internal and external fields generates complex resonance structures that redistribute energy across a broader spectrum. While the exact mode requirement to reach a compression threshold (e.g., 99% energy) is material-dependent, the complexity of the spectral envelope consistently scales with the size parameter until reaching the geometric scattering region, where amplitudes eventually plateau or gradually decrease ($10^{-3}$ to $10^{1}$), as the fine-scale resonance features transition into a more stable, slowly varying extinction efficiency.

To quantify this energy redistribution, we utilize a power-law reference $1/m^\beta$. In harmonic analysis, the decay rate $\beta$ correlates to signal smoothness: a decay of $\beta = 2$ implies a continuous signal with sharp slope discontinuities (e.g., high-$Q$ resonances), while faster decays ($\beta > 2$) indicate higher differentiability and spectral smoothness. We observe that, for all polymers, small spheres exhibit rapid decay ($\beta > 2$), reflecting the highly compressible and smooth nature of the Rayleigh envelope. Conversely, the onset of sharp, interference-driven Mie resonances at $r \geq 0.1\ \mu m$ introduces corners and steep gradients in the spectral data, pushing the decay closer to the $m^{-2}$ limit. As $r$ increases, the energy distribution in the spectrum flattens—evidenced by a flattening of the decay slope—as energy spreads into high-frequency components corresponding to these fine-scale oscillations.

The superior energy compaction of the DCT is visually confirmed by the rapid decay of its coefficients compared to those of the FFT across all scattering regimes (see **Figures S7–S12**). While the FFT spectra exhibit a sustained high-frequency floor—a direct indicator of spectral leakage necessitated by non-periodic boundary conditions—the DCT effectively concentrates the physical information into a sparse, low-dimensional subset of modes. This accelerated decay rate underscores the DCT's ability to represent the extinction manifold using significantly fewer degrees of freedom without compromising morphological detail.

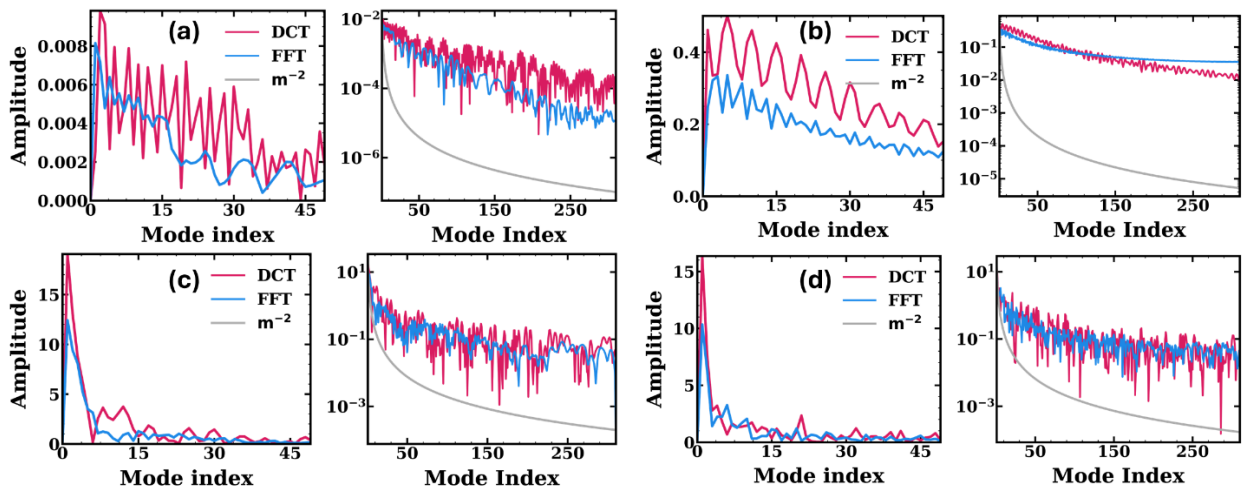

Figure 3: Spectral Energy Distribution and Energy Compaction. Comparison of the amplitude spectra for DCT and FFT bases across representative PC microsphere radii: **(a)** $r = 0.01\ \mu m$, **(b)** $r = 0.1\ \mu m$, **(c)** $r = 1.0\ \mu m$, and **(d)** $r = 10.0\ \mu m$. **(Left Panels):** The first 50 harmonic modes on a linear scale, demonstrating that the DCT concentrates the vast majority of signal energy into the lowest-order coefficients. **(Right Panels):** Semilog representation of the full spectral range, illustrating the coefficient decay of the DCT and FFT compared to the $m^{-2}$ limit, corresponding to continuous signals with sharp discontinuous.

The core of our information-theoretic analysis is captured in **Figure 4**, which depicts the normalized spectral entropy (**H**) and sparsity requirements of PC microspheres across the Rayleigh, Mie, and Geometric scattering regimes. While the library contains six distinct organic polymers, we present PC as a representative dielectric due to its hybrid aliphatic-aromatic structural complexity. Comprehensive results for the remaining five polymers, confirming the material-independence of our framework, are provided in the Supplementary Information **Figure S13** and **Tables S1-S6.**

For all materials, $H$ peaks within the Mie transition region ($r \approx 0.1\ \mu m$), signifying a global maximum in structural complexity (**Figure 4a**). This is the information bottleneck—the point where the extinction manifold requires the highest number of degrees of freedom for representation. Notably, a divergence in entropy behavior occurs between the two transforms. **1)** The DCT consistently exhibits a lower entropy and superior energy compression. An apparent entropy rise in the Rayleigh regime for certain polymers is a numerical artifact where the physical signal ($10^{-5}$ to $10^{-2}$) nears the noise floor, forcing the inclusion of stochastic noise modes. As the physical signal rises in the Mie regime, the DCT's superior compaction becomes dominant. **2)** The FFT's higher entropy is a "Spectral Leakage Contribution" (**Figure 4a**), arising from the periodic extension's failure to reconcile the non-periodic boundaries of physical extinction data.

Significantly, we find that standard Hann windowing—the conventional solution for leakage—is counterproductive for $Q_{ext}$ reconstruction (**Figure S14**). While windowing suppresses leakage artifacts in the transform domain, the inverse process recovers a "windowed" version of the physical signal. Attempting to recover the original $Q_{ext}$ by dividing by the window function introduces extreme numerical instability at the boundaries where the window approaches zero. Thus, the DCT is not merely a mathematical preference but a physical necessity for preserving boundary-encoded scattering information.

Our benchmarks confirm three primary conclusions: **1)** spectral complexity of the $Q_{ext}$ signal is non-monotonic, peaking at the onset of the Mie regime ($r \approx 0.1\ \mu m$) and stabilizing shortly after; **2)** regardless of the specific polymer, $Q_{ext}$ remains exceptionally sparse (and hence, compressible) in the DCT basis; and **3)** the DCT efficiency follows a U-shaped curve, performing best for very small ($r < 0.1\ \mu m$) and very large ($r \geq 10\ \mu m$) particles, where signals are either highly sparse or systematically structured, while the information challenge is concentrated in the critical transition regime ($r \approx 0.1\ \mu m$), where the primary interference maxima emerge.

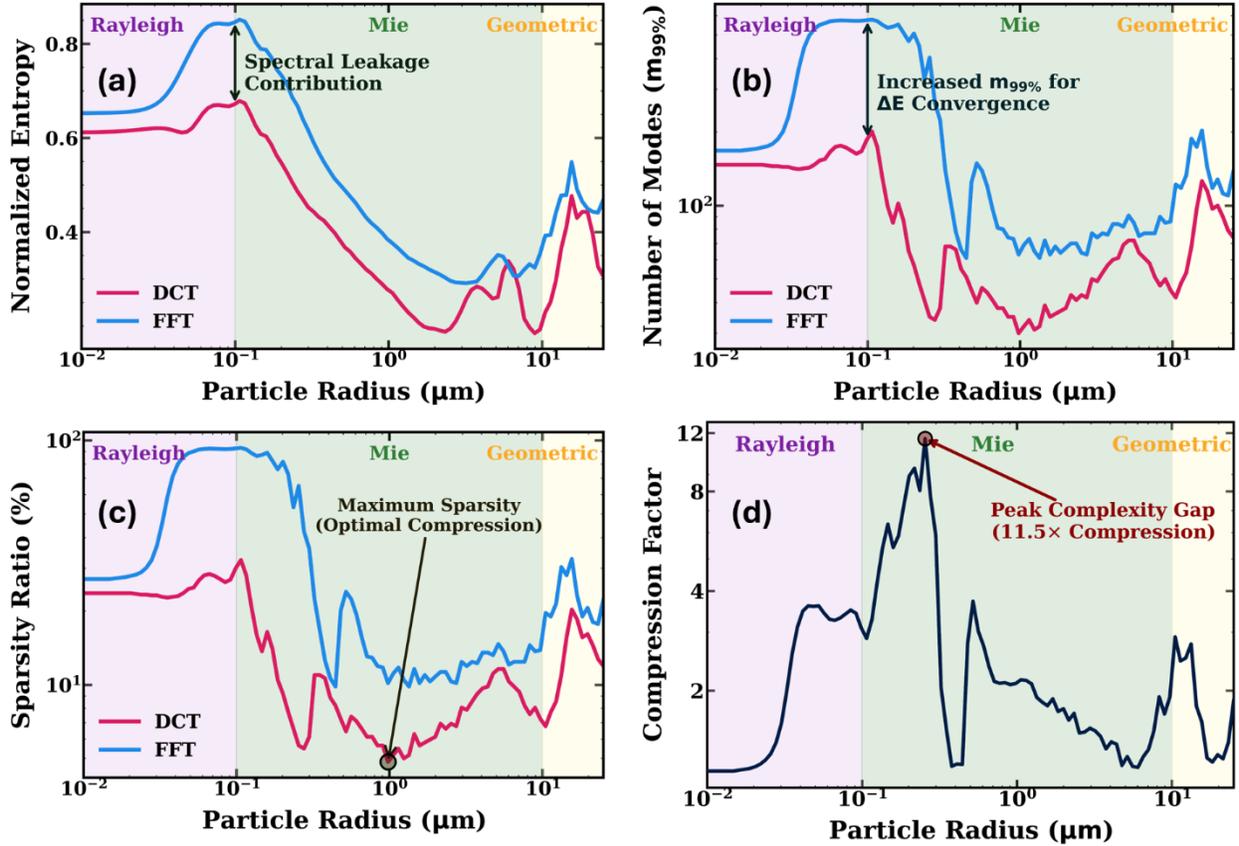

**Figure 4: Information Complexity and Spectral Sparsity of the $Q_{ext}$ Manifold.** Comprehensive analysis of the encoding efficiency for Polycarbonate (PC) across the Rayleigh, Mie, and Geometric scattering regimes ($r = 0.01$ to $25 \, \mu m$). **(a) Evolution of Spectral Entropy.** Normalized spectral entropy as a function of particle radius, illustrating the transition from the low-entropy Rayleigh regime to the peak complexity of the Mie transition ($r \approx 0.1 \, \mu m$). The consistently lower entropy profile of the DCT demonstrates its superior capacity for information compaction. **(b) Sparsity Analysis and Modal Requirements.** The number of coefficients ($m_{99\%}$) required to retain 99% of the cumulative spectral energy. The DCT consistently outperforms the FFT, particularly at the Mie transition ($r \approx 0.1 \, \mu m$). Note: The apparent increase in required modes for $r \leq 0.03 \, \mu m$ is a numerical artifact caused by the vanishingly small amplitude ($\sim 10^{-5}$ to $10^{-2}$) in the Rayleigh limit, where stochastic numerical noise dominates the cumulative sum. **(c) Spectral Sparsity Ratio.** The percentage of total coefficients required to reach the 99% energy threshold. The relative efficiency of the transforms is strictly governed by the underlying scattering regime; signal compressibility reaches a global minimum at the information bottleneck ($r \approx 0.1 \, \mu m$). The DCT consistently outperforms the FFT across the entire manifold, achieving maximum sparsity and optimal compression as the system approaches the Geometric region ($r \approx 10 \, \mu m$), where the FFT's periodic assumptions lead to increasing modal redundancy. **(d) Comparative Basis Compaction Analysis.** Ratio of FFT to DCT mode requirements for 99% energy capture across different scattering regimes. The plot demonstrates a significant reduction ($\sim 12 \times$) in representation dimensionality when using the DCT, particularly for particles at the onset of the Mie resonance size range.

## Performance under Cumulative Energy Threshold Truncation

To quantify the intrinsic dimensionality of the extinction manifold, we determined the minimal mode count ($m_{required}$) required to retain 90%, 95%, and 99% of the cumulative spectral energy (**Figure 4b**, **Figures S15-S20**, and **Tables S7-S12**). Theoretically, $m_{required}$ should scale with signal complexity, reaching its global maximum at the onset of the Mie transition regime ($r \approx 0.1\ \mu m$). While materials like PC, PDMS, and PS follow this trajectory, others (e.g., PMMA, PEI, and PET) show an apparent sparsity failure in the Rayleigh limit ($r < 0.1\ \mu m$), where $m_{required}$ artificially inflates. As established previously, this is a consequence of the numerical noise floor: when the physical signal vanishes, energy-based thresholds force the DCT to aggregate hundreds of stochastic noise modes.

In the Mie transition regime ($r \approx 0.1\ \mu m$), where the physical signal dominates the noise, DCT's superior energy compaction is evident. For example, in PC microspheres, the DCT captures complex resonances using significantly fewer modes ($m_{90\%} = 66, m_{95\%} = 94, m_{99\%} = 189$) than the FFT ($m_{90\%} = 281, m_{95\%} = 418, m_{99\%} = 576$) across all energy thresholds. This efficiency is further quantified by the spectral ratio ($\frac{m_{99\%}}{N} * 100\%$) in **Figure 4c** (and **Figure S21** for the remaining polymers), which confirms that the DCT maintains higher information density per coefficient at the prescribed energy level. These results further identify the scattering regions of maximum sparsity, representing the optimal compression regime for a material's $Q_{ext}$ signal. Crucially, these regions define the 'information-theoretic limit' of the system, where a minimal set of spectral measurements is sufficient to reconstruct the total extinction profile.

The most rigorous evidence for this bottleneck is the modal compression ratio ($m_{FFT}/m_{DCT}$) presented in **Figure 4d** (and **Figure S22** for the remaining five polymers). For PC microspheres at the 99% energy threshold, this ratio is minimized in the Rayleigh and Geometric regimes, where signals are either inherently sparse or exhibit high-frequency stability, with the FFT requiring only $\sim 2.2 \pm 1.1$ and $\sim 1.9 \pm 0.6$ times more modes than the DCT, respectively. Conversely, the complexity gap reaches a dramatic maximum of $\sim 11.5$ times within the Mie transition regime ($r \approx 0.1\ \mu m$). Across our six-polymer library, this complexity gap remained consistently high, averaging a staggering $\sim 12 \times$ increase. While the high compression factor observed in the PC Rayleigh regime is partially an artifact of the stochastic noise floor—forcing the DCT to aggregate non-physical noise modes—the multi-fold increase in FFT modes required at the Mie onset is a direct consequence of the non-periodic, overlapping resonances characterizing this regime. In this region, spectral leakage and boundary artifacts inherent to the Fourier basis are maximized; in contrast, the DCT's boundary-matching symmetry maintains high efficiency. This divergence in the compression gap unambiguously identifies the Rayleigh-Mie transition ($r \approx 0.1\ \mu m$) as the point of maximum information-theoretic complexity, where the physical scattering process is mathematically most difficult to resolve.

As summarized in **Table S19**, this efficiency advantage is not static but scales proportionally with the demand for fidelity. The peak ratio in the transition regime expands from $\sim 4.6 \times$ at 90% energy to $\sim 12 \times$ at 99% energy. This trend identifies a precision-complexity-paradox inherent to Fourier-based spectroscopy: the more physical detail one attempts to resolve, the more the FFT obscures it with mathematical artifacts. To capture the final 1% of spectral energy, the FFT must aggregate a disproportionately large number of high-frequency modes to compensate for persistent boundary leakage This phenomenon represents the FFT mathematically fighting its own periodic constraints, as evidenced by the asymptotic convergence of the cumulative energy curves (**Figures S23-S28**). In contrast, the DCT's intrinsic symmetry resolves these delicate resonance tails with minimal modal expansion, suggesting it is increasingly indispensable as the demand for spectroscopic precision grows.

While the information bottleneck is a persistent feature across all analyzed polymers, its precise morphological coordinate ($r$) exhibits slight inter-species variation, typically ranging from $0.15\ \mu m$ to $0.35\ \mu m$ (**Table S19**). We attribute this variation to the material-specific complex refractive index ($\tilde{n} = \tilde{n}' + i\tilde{n}''$), governs internal phase shifts and resonance damping. Specifically, the real part ($\tilde{n}'$) dictates the optical path length and resonance frequency, while the imaginary part ($\tilde{n}''$) modulates the amplitude of interference ripples. This interplay determines the exact onset of peak spectral complexity, suggesting that while the existence of the bottleneck is a universal geometric property of Mie scattering, its precise location is modulated by the material's dielectric function.

Furthermore, we find that the observed coordinate of the bottleneck is coupled to both the choice of basis and the energy retention threshold. As the fidelity requirement increases from 90% to 99%, the bottleneck shifts and sharpens (**Table S19**). This phenomenon arises because lower fidelity thresholds primarily capture the broad, low-frequency scattering envelope, whereas high-fidelity constraints force the basis to resolve the high-frequency, interference-driven resonances that characterize the transition regime.

Consequently, the intrinsic dimensionality of the Mie manifold is not a static value, but a dynamic, threshold-dependent property. Interestingly, as the fidelity requirement increases from 90% to 99%, the bottleneck coordinate exhibits a non-linear shift—initially sharpening at smaller radii ($r \approx 0.15\ \mu m$) before migrating toward a larger radius ($r \approx 0.26\ \mu m$) at the 99% threshold. This rebound at the highest energy retention level suggests a hierarchy of complexity: while the primary scattering envelope reaches its maximum disorder early in the transition, the fine-scale Mie ripples—which are only resolved at the 99% threshold—continue to accumulate structural complexity as the particle size increases. This demonstrates that the worst-case complexity is not only revealed under high-fidelity constraints but is physically localized within a specific sub-regime of the Mie transition where interference effects are most mature.

Despite these minor numerical shifts, the spatial coincidence of the complexity peak across both DCT and FFT frameworks—and its invariance under noise—confirms that the bottleneck is a fundamental physical property of the Mie manifold rather than a mathematical artifact. We adopt the DCT-derived peak as our reference coordinate because its superior energy compaction provides the most transparent "window" into the manifold's intrinsic dimensionality. A systematic study of the correlation between this bottleneck coordinate and the integrated refractive index over the mid-IR band may further refine the predictive mapping of spectral sparsity for future instrument design.

**Figure 5** illustrates the sparse reconstruction of $Q_{ext}$ for PC using a 99% energy threshold. For a comprehensive visualization of these reconstructions across all three energy thresholds (90%, 95%, and 99%) and for the full six-polymer library, the reader is referred to **Figures S29–S46**. Collectively, these results quantify the fundamental trade-off between modal compression and morphological fidelity, demonstrating that the DCT basis maintains superior structural accuracy even at high compression ratios. As expected, the quality of the $Q_{ext}$ reconstruction improves monotonically as the energy threshold increases from 90% to 99%. In the 90% and 95% cases, significant data compression is achieved at the expense of high-frequency detail. Notably, while the lower thresholds lead to the loss of fine-scale fringes and secondary oscillations, the overall global shape and magnitude of the $Q_{ext}$ curve remain preserved. This suggests that the primary resonance peaks are contained within the first few percent of the transform's energy spectrum, while the oscillatory ripple structure constitutes the lower-energy spectral tail.

The fidelity of the reconstruction exhibits a strong dependency on the particle radius. In the Rayleigh regime ($r < 0.1\ \mu m$), the smooth $Q_{ext}$ profiles allow for near-perfect reconstruction. Even at a 90% energy threshold (**Figures S41-S46**), both the major peaks and the majority of small-scale oscillations are accurately captured. However, as size increases, the quality of the reconstruction progressively decreases. Larger radii introduce highly oscillatory behavior into the $Q_{ext}$ signal. Because these oscillations are spread across a broader range of frequencies, a fixed energy threshold necessarily discards more of the fringe detail, resulting in a smoothed approximation of the complex Mie efficiency factors. This shift in reconstruction behavior highlights that error is not uniformly distributed across the parameter space. We find that while reconstruction fidelity scales predictably with the energy retention threshold ($90\% - 99\%$), the error morphology —whether it manifests as smooth approximations or high-frequency artifacts— is fundamentally dictated by the physical scattering regime rather than the choice of basis.

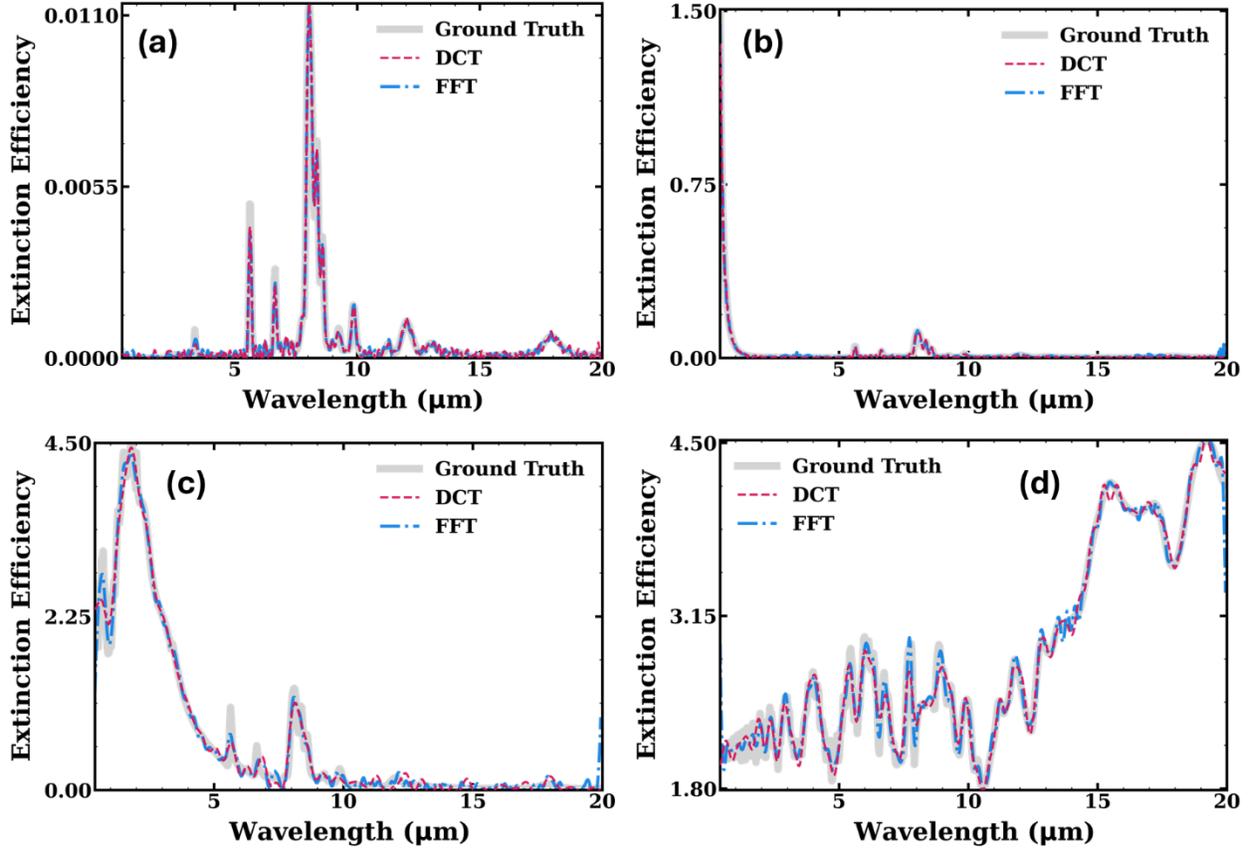

**Figure 5: Fidelity of Sparse Reconstruction at the 99% Energy Threshold.** Comparative reconstruction of $Q_{ext}$ profiles of PC microspheres using a truncated subset of DCT and FFT coefficients corresponding to the 99% cumulative energy threshold. Profiles are shown for particle radii spanning the Rayleigh to Geometric scattering regimes: **(a)** $r = 0.01\ \mu m$, **(b)** $r = 0.1\ \mu m$, **(c)** $r = 1.0\ \mu m$, and **(d)** $r = 10\ \mu m$. While both bases achieve comparable reconstruction fidelity for a fixed energy threshold, the DCT provides a more computationally efficient mapping of the extinction manifold. Unlike the FFT, which requires a broader modal distribution to compensate for boundary-induced spectral leakage, the DCT concentrates the signal's essential physical information into a significantly more compact subset of coefficients. This visual validation confirms that the DCT effectively captures the essential physical information of the extinction manifold within a highly compressed modal subset, providing a more streamlined basis for hardware-limited spectroscopic applications.

Our comprehensive error analysis (**Figure 6** and **Figures S47–S64**) confirms that the onset of the Mie transition ($r \approx 0.1\ \mu m$) represents a complexity bottleneck, evidenced by a prominent error plateau across all metrics (RMSE, $E_{max}$, and LRMSE), where high-frequency resonance ripples challenge truncated basis sets. Notably, while the FFT may occasionally exhibit a lower LRMSE at the immediate onset of the transition, it does so only through extreme over-parameterization, utilizing a significantly larger basis set to meet the fixed energy threshold. For instance, to reach a target LRMSE of $\sim 10^{-1}$ at $r \approx 0.1\ \mu m$, the DCT requires $\sim 92\%$ fewer modes than the FFT. Furthermore, the DCT maintains a more monotonic and stable error path, providing a more reliable "worst-case" bound ($E_{max}$). This suggests that the DCT is indispensable for high-precision applications, such as atmospheric sensing, where pointwise divergence at specific resonance peaks must be minimized.

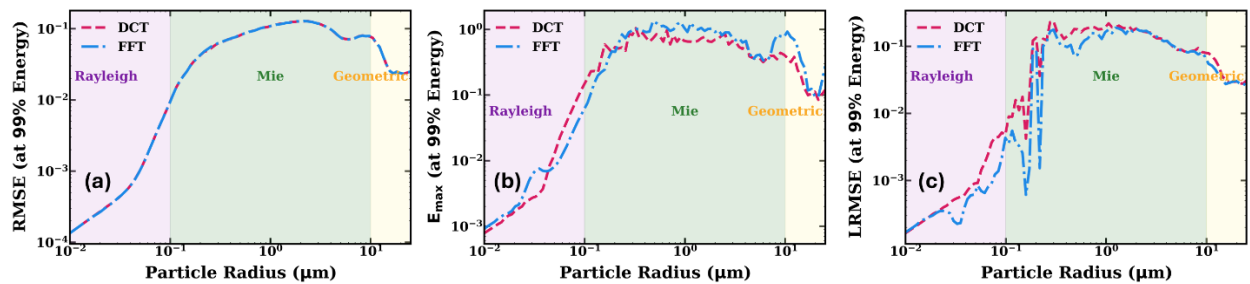

**Figure 6: Reconstruction Error Analysis for PC Microspheres at the $99\%$ Energy Threshold. (a) Global RMSE** as a function of particle radius, illustrating the overall convergence stability of the DCT and FFT bases. **(b) Maximum Absolute Error ($E_{max}$)**, highlighting the "worst-case" pointwise divergence typically located at sharp resonance peaks. **(c) Local RMSE (LRMSE)**, resolving the localized error dynamics within specific scattering regimes. Note the prominent error peak in the Mie transition regime ($r \approx 0.1\ \mu m$) across all metrics, identifying the region of maximum information-theoretic complexity where non-periodic resonance oscillations are most difficult to resolve.

## Performance under Sequential Mode Truncation (Fixed Budget)

To evaluate compaction efficiency independently of the noise floor, we performed sequential truncation using a fixed mode budget $M$. **Figure 7** illustrates the energy fractions ($E_5$, $E_{10}$, and $E_{20}$) captured within the top $M = 5$, 10, and 20 modes. Supplementary data for the remaining five polymers and comprehensive raw data are available in **Figures S65–S69** and **Tables S13–S18**, respectively. Both transforms exhibit a characteristic "dip and rise" trend in energy compactness, where efficiency reaches a global minimum at onset of the Mie transition ($r \approx 0.1\ \mu m$).

This regime represents a physical information bottleneck: a sensor limited to $M$ modes capture its lowest possible fraction of total signal power here. However, even at this bottleneck, the DCT maintains a significant advantage. While the FFT distributes its information budget into boundary artifacts—essentially wasting coefficients on the Gibbs-like effect of periodic extension—the DCT's superior boundary handling preserves signal energy within a minimal subset of dominant modes.

Once the system bypasses the information bottleneck ($r \approx 0.1\ \mu m$), both transforms show accelerated energy convergence toward the 100% limit. This has profound implications for the effective Signal-to-Noise Ratio (SNR), especially in terms of compressibility and noise suppression. In the post-transition and Geometric regimes, both bases concentrate $> 95\%$ of spectral energy within the first $M = 20$ modes. Consequently, reconstruction can be achieved using a highly restricted integration bandwidth. By discarding high-frequency coefficients—where the FFT typically manifests a noise floor due to leakage—the framework maximizes the information-per-sensor.

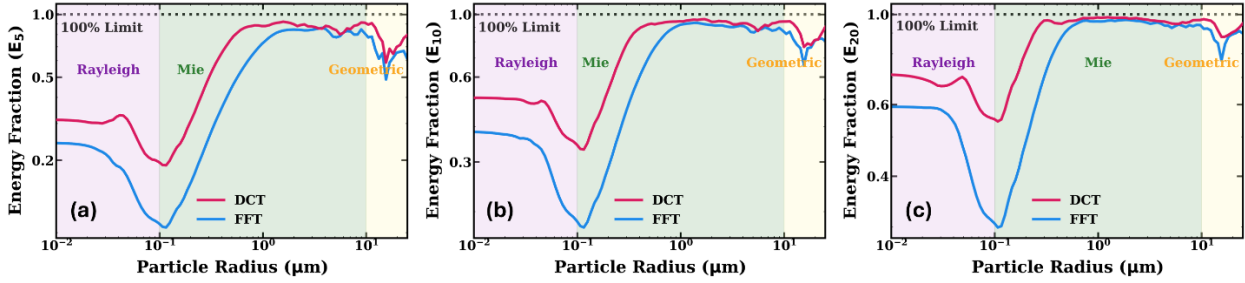

Figure 7: **Comparative Analysis of Spectral Energy Concentration.** Energy fraction in PC microspheres captured within the first $M$ modes: **(a)** $M = 5$, **(b)** $M = 10$, and **(c)** $M = 20$, for DCT and FFT-approximated $Q_{ext}$ signals across the full range of particle radii ($r = 0.01$ to $25\ \mu m$). The plots quantify the low-rank sparsity of the extinction manifold; the DCT consistently concentrates a higher proportion of spectral energy in fewer coefficients than the FFT, particularly at the onset of the Mie transition regime ($r \approx 0.1\ \mu m$). This accelerated energy convergence provides the physical basis for high-fidelity reconstruction using extremely sparse sensor arrays.

The four-metric validation in **Figure 8** demonstrates that the DCT's advantage is not merely statistical (RMSE) but fundamentally structural (LRMSE and $E_{max}$). The evolution of global RMSE as a function of retained modes ($M$) provides a definitive measure of energy compaction (**Figure 8a** and **Figure S70**), revealing that the DCT consistently achieves a steeper convergence gradient and a lower error floor than the FFT.

Beyond simple compaction, the DCT demonstrates superior pointwise reconstruction accuracy that scales non-linearly with the number of modes. At the critical Mie transition, this performance gap reaches its zenith: for a fixed truncation of only $M = 20$ modes, the DCT achieves a reconstruction error nearly an order of magnitude lower than the FFT. This $6.9\times$ increase in precision confirms that the DCT is not merely a compact representation of the dielectric extinction spectrum, but a more physically resonant basis for resolving fine-structured oscillations.

Notably, this maximum performance gap occurs at low modal counts ($M = 20$) rather than at the highest resolutions ($M = 100$). This reveals a high-fidelity shortcut: while the FFT can eventually brute-force a reconstruction using a massive basis set, the DCT captures the essential physics of the Mie transition far more rapidly. For real-time applications such as high-throughput clinical cytology or satellite-based remote sensing, this rapid convergence translates directly into reduced memory overhead and lower latency.

While RMSE tracks global energy loss, the maximum absolute error ($E_{max}$) and local RMSE (LRMSE) identify worst-case localized failures (**Figures 8b**, **8c**, and **S71-S72**). For a fixed mode budget, the DCT maintains a lower and more stable $E_{max}$ ceiling, particularly in the oscillatory regime, effectively stress testing the basis functions at steep resonance gradients. While the FFT shows a marginal local advantage in isolated regions—likely due to its complex basis handling unstructured oscillations—the DCT rapidly regains dominance as particle size increases. This localized stability is critical for preserving the morphological precision of Mie resonances; as these peaks serve as the spectral fingerprints required for high-confidence material characterization, the DCT ensures that diagnostic features are not smoothed over or lost to Fourier artifacts.

A critical insight emerges from **Figure 8d** (and **Figure S73**), which plots RMSE as a function of the retained energy fraction. Across various particle sizes (indicated by color), the DCT consistently initiates at a higher energy fraction for low mode budgets ($M$), indicating superior information retention for a lower experimental cost. This trend is most prominent outside the Mie transition regime. Geometric-regime particles exhibit the highest compressibility, demonstrating the greatest energy retention for the smallest number of modes. Furthermore, regardless of particle size, the DCT approaches the theoretical limit (100% energy retention) significantly faster than the FFT.

Crucially, **Figure 8d** demonstrates that energy retention alone is an insufficient metric for reconstruction quality. For example, at $r = 0.1\ \mu m$, the first 10 DCT modes capture $\sim 37\%$ of the signal energy, compared to only $\sim 19\%$ for the FFT. Even when both bases are tuned to retain the same energy fraction, the DCT achieves higher fidelity using fewer modes. This proves that the energy accumulated by the DCT is intrinsically more relevant to the signal's physical geometry. The FFT's energy is "diluted" by spectral leakage; it must aggregate high-frequency terms simply to cancel out its own periodic artifacts, which does not contribute to feature reconstruction. In contrast, the DCT's energy is mapped directly to the underlying physical morphology, making it the indispensable basis for low-loss representation of Mie scattering spectra.

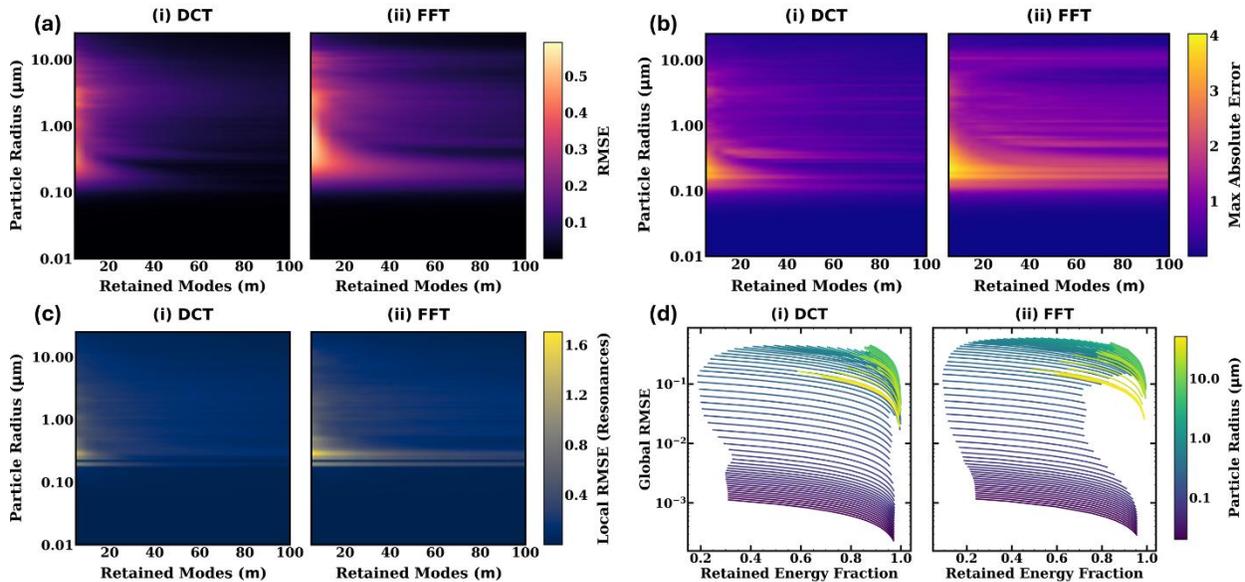

Figure 8: **Multiscale Error Analysis and Convergence Benchmarking for PC microspheres.** Comparative performance of DCT and FFT bases across Rayleigh, Mie, and Geometric scattering regimes ($r = 0.01$ to $25\ \mu m$). **(a) Global Convergence of Reconstruction Error.** Root-Mean-Square Error (RMSE) as a function of retained modes ($M$). The DCT demonstrates a superior convergence rate, maintaining a lower error floor than the FFT at most truncation levels. This is most pronounced at the information bottleneck ($r \approx 0.1\ \mu m$), where the DCT's handling of non-periodic features significantly reduces required modal cardinality. **(b) Peak Error and Artifact Suppression.** Maximum absolute error ($E_{max}$) versus $M$. The DCT consistently suppresses localized artifacts, such as the Gibbs phenomenon, preserving the structural integrity of sharp resonance features and preventing non-physical spectral distortions. **(c) Localized Fidelity at**

**Resonance Maxima.** Localized RMSE (LRMSE) calculated within the spectral neighborhoods of primary resonance peaks. By isolating these high-gradient features, the plot confirms the DCT's ability to resolve sharp physical transitions without the spectral blurring characteristic of Fourier-truncated approximations. **(d) Error-Energy Efficiency.** Reconstruction RMSE as a function of retained spectral energy fraction. The DCT achieves lower error at equivalent energy levels, demonstrating that it is more informationally efficient; whereas FFT energy is often "diluted" by spectral leakage, DCT energy is directly mapped to the underlying physical morphology of the $Q_{ext}$ signal.

## Noise Sensitivity and the Resilience of the Information Bottleneck

To evaluate the robustness of the inverse transformation framework, we subjected the extinction manifold to $1-10\%$ additive Gaussian noise. This stress-test (**Figures 9** and **S74–S78**) serves as a representative model for real-world experimental stochasticity, proving that the identified information-theoretic constraints are not fragile numerical artifacts, but persistent features of the scattering physics.

**Figure 9a** presents the full-rank reconstruction (without coefficient truncation) of a noisy $Q_{ext}$ spectrum for a spherical particle of radius $r = 1.0\ \mu m$. By maintaining the full set of coefficients, this baseline comparison isolates the inherent stability of the transformation bases under stochastic perturbations, ensuring that observed errors are attributable to noise propagation rather than information loss through sparsification.

The normalized entropy heatmaps (**Figure 9b**) provide a visual proxy for the information content across the scattering regimes. Crucially, the information bottleneck at the Mie transition ($r \approx 0.1\ \mu m$) remains spatially and structurally invariant even as the noise floor rises. While the Rayleigh region shows an artificial entropy spike due to vanishingly low signal-to-noise ratios, and the Geometric region shows noise-induced disorder, the bottleneck consistently represents the global maximum of structural complexity. This invariance confirms that the bottleneck is a fundamental topological feature of the Mie manifold; noise does not shift the location of maximum complexity, it simply exaggerates the challenge of resolving it. Additionally, across all radii and noise levels, the DCT maintains consistently lower entropy than the FFT. This gap narrows as noise increases, as the Gaussian white noise—characterized by maximum entropy—eventually dominates the structured spectral coefficients of the scattering signal.

Profiling the global RMSE, maximum absolute error, and local RMSE at 0% and 10% noise levels demonstrates the robustness of both transforms (**Figure 9c**). Interestingly, in certain regimes, the reconstruction error at 10% noise is lower than in the clean baseline; This phenomenon likely occurs because low-energy modes fall below the noise floor, effectively pruning high-frequency jitters and reducing the reconstructed signal complexity. While both transforms yield similar absolute error margins across the radius sweep, their underlying efficiency in achieving this accuracy differs fundamentally.

Despite similar error profiles (**Figure 9c**), the mode budget required to satisfy the 99% energy threshold reveals the DCT's persistent advantage (**Figure 9d**). The FFT requires significantly more modes than the DCT to capture the same information, particularly at the Mie transition. This is quantified by the compression factor ($m_{FFT}/m_{DCT}$) in **Figure 9e**.

As illustrated in **Figure 9e**, the DCT demonstrates a superior structural alignment with the high-frequency oscillations of the Mie manifold, achieving a $12\times$ efficiency advantage in mode requirements compared to the FFT at the $99\%$ energy threshold. However, as noise increases, the compression factor decreases, and the landscape becomes more uniform. This trend is driven by two competing mechanisms. Firstly, Gaussian noise is spectrally flat, distributing energy across all frequencies and forcing both transforms to allocate modes to stochastic fluctuations. Secondly, at high noise levels, the global noise floor rises above the boundary errors that typically cripple the FFT. This masks the DCT's boundary-handling advantage, making the signal appear as an uncompressible, high-entropy cloud.

Comparing performance envelopes under $10\%$ noise (**Figure 9f**) reveals that the DCT maintains a superior efficiency ceiling. For a fixed mode budget, the addition of noise reduces the total retained energy, yet the DCT minimizes the resulting error more effectively than the FFT across the entire radius sweep.

This confirms that the DCT's energy compaction is not merely an "ideal case" benefit. Instead, it proves that the DCT coefficients align with the true physical signal of the Mie manifold, while the FFT's reliance on high-frequency leakage modes makes it fundamentally hypersensitive to noise. This resilience ensures that the transition from theoretical modeling to real-world hardware—where noise floors are a persistent constraint—does not compromise the structural integrity of the reconstruction.

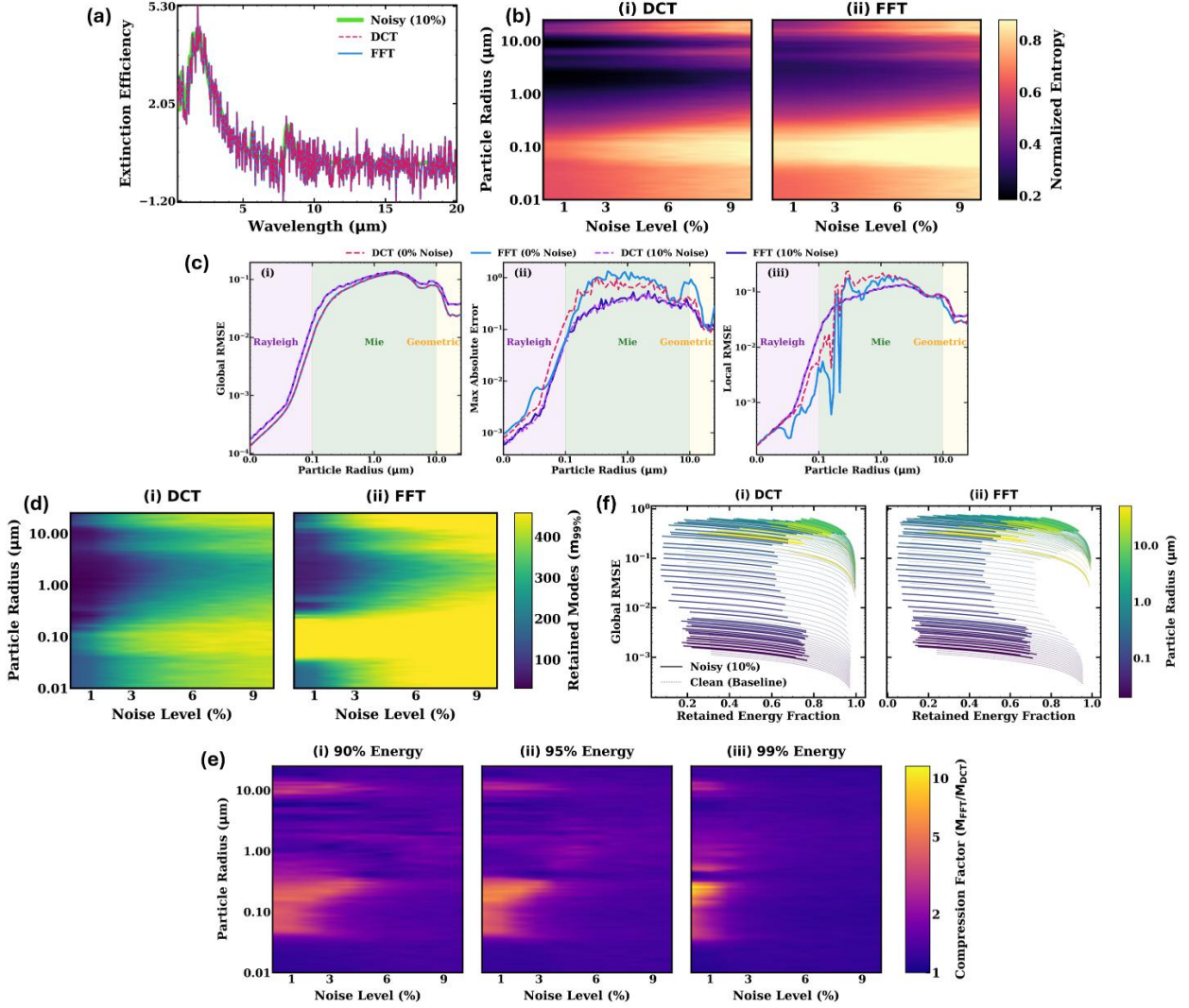

**Figure 9: Resilience and Invariance of the Information Bottleneck under Stochastic Perturbation.** Comprehensive noise sensitivity analysis for PC microspheres, demonstrating that the structural complexity peak remains a fundamental topological feature of the $Q_{ext}$ manifold regardless of experimental noise. **(a) Full-Rank Reconstruction under 10% Noise:** Noisy $Q_{ext}$ spectrum ($r = 1.0\ \mu m$) reconstructed without truncation. This baseline confirms the numerical stability of both bases, isolating noise propagation from sparsification-induced error. **(b) Entropy and Information Invariance:** Normalized entropy maps for **(i)** DCT and **(ii)** FFT as a function of noise level and radius. Crucially, the peak complexity—the information bottleneck—remains spatially locked at the Mie transition ($r \approx 0.1\ \mu m$), proving its invariance to noise. While the entropy gap narrows as white noise masks the structured signal, the bottleneck's position remains a fixed coordinate of the scattering physics. **(c) Multi-Metric Error Profiling:** Radius sweep showing **(i)** Global RMSE, **(ii)** Maximum Absolute Error, and **(iii)** Local RMSE for both clean (0%) and noisy (10%) datasets. While noise elevates the error floor, the relative performance margins and the characteristic error plateau at the Mie transition remain structurally consistent. **(d) Mode Retention Dynamics ($m_{99\%}$):** Heatmaps for **(i)** DCT and **(ii)** FFT illustrating the number of modes required for 99% energy retention. The FFT requires significantly higher modal cardinality

(brighter regions) in the Mie transition zone ($r \approx 0.1 \ \mu m$) compared to the DCT, a disparity that is exaggerated rather than eliminated by increasing noise. **(e) Compression Efficiency ($m_{FFT}/m_{DCT}$):** Heatmaps showing the compression advantage at **(i)** $90\%$, **(ii)** $95\%$, and **(iii)** $99\%$ energy thresholds. The DCT's efficiency advantage is most resilient in the Mie transition ($r \approx 0.1 \ \mu m$); the narrowing of this factor at high noise represents the stochastic masking of the structured scattering signal. **(f) Performance Envelopes at $10\%$ Noise:** Global RMSE vs. retained energy fraction for **(i)** DCT and **(ii)** FFT. The solid-colored lines represent the noisy performance across the radius sweep, while the light-grey dashed lines establish the $0\%$ noise baseline floor. The DCT maintains a superior envelope even under maximum noise. Note the vertical shift and convergence truncation in the Mie regime (yellow/green), where the noise floor prevents the deep RMSE drops seen in the Rayleigh regime (purple), further identifying the transition region as the ultimate limit of signal compressibility.

## Hardware Co-Design: Optimal Sensor Placement and Reconstruction Stability

The ultimate utility of the identified information bottleneck is the ability to transition from high-resolution computational models to "thin" experimental hardware. By exploiting the sparsity of the $Q_{ext}$ manifold in the DCT domain, we demonstrate a framework for optimal sensor placement that achieves significant reduction in hardware complexity while maintaining high-fidelity reconstruction.

To translate theoretical sparsity into physical hardware, we utilize a sensitivity-based heuristic to identify the spectral coordinates that maximize information gain. This methodology isolates the rows within the DCT basis matrix that concentrate the highest energy for the dielectric signal class. As illustrated in **Figure 10a**, the DCT-derived sensitivity score acts as a structural proxy for the underlying Mie scattering physics. The peaks in this score (solid navy blue) naturally align with regions of high information density, defined by the physical curvature ($|\frac{\partial^2 Q_{ext}}{\partial \lambda^2}|$) (dashed orange) of the extinction profile (solid teal). This creates a gradient-weighted sampling mask that targets regions of rapid signal change.

Unlike the FFT (dotted gray in **Figure 10a**), which generates a non-physical "barcode" sensitivity, that fails to distinguish between high-curvature resonances and redundant flat regions, the DCT concentrates sensors at the spectral boundaries ($\sim 2.5 \ \mu m$ and $20 \ \mu m$) and at internal resonance inflection points. This dual-scale sampling anchors the global envelope while resolving localized high-gradient spikes, enabling a 12-fold reduction in data density compared to standard FFT-based approaches.

The proposed framework establishes a fundamental lower bound on hardware complexity, which we define relative to a typical high-resolution Nyquist baseline ($N_{high}^{(\lambda)} = 350$). This baseline corresponds to a standard spectral resolution of $10 \ cm^{-1}$ across the mid-infrared band ($2.5 - 20 \ \mu m$)—the minimum density required to resolve secondary Mie resonances in organic dielectric polymers. While our internal simulations utilize a higher-density oversampled grid ($N = $

620) to ensure numerical convergence, the hardware reduction efficiency is benchmarked against this $N_{high}^{(\lambda)} = 350$ standard to reflect real-world spectroscopic requirements.

To evaluate the performance of our compressed sensing architecture under worst-case complexity, we anchor the hardware configuration at $r \approx 0.1\ \mu m$, the representative locus of the Mie transition bottleneck. While the peak complexity coordinate varies slightly with material species and fidelity thresholds, this coordinate serves as a rigorous baseline for assessing the DCT's ability to resolve critical interference features with minimal sensor overhead. In this region, where spectral entropy peaks and information density is most dispersed, the system achieves reconstruction of over 95% of the spectral energy using only $P = 170$ sensors—a 51.4% reduction in hardware complexity. For particles in the Rayleigh or Geometric regimes, where the signal manifold is inherently sparser, this requirement drops even further to as few as $P = 22$ sensors, representing hardware reductions of 93.7%. By operating near the theoretical cardinality bound ($P \approx 2M$), the system effectively eliminates the massive redundancy inherent in conventional spectroscopy, enabling "thin" sensing architectures without sacrificing morphological precision.

To ensure the numerical stability of the inverse problem, we analyzed the condition number ($\kappa$) of the sensing matrix ($\Theta = \Phi\Psi$) with respect to the $l_2$ norm [9]. Physically, $\kappa$ represents the error amplification factor; a high $\kappa$ indicates that minor detector noise could induce non-physical oscillations.

While the initial configuration ($P = 20$ sensors, $M = 10$ modes) is physically optimal for anchoring boundaries (**Figure 10b**), it results in an ill-conditioned system ($\kappa \approx 10^3$). By performing a global stability sweep (**Figure 10c**), we identified a stability plateau at $M = 100$ and $P = 204$. This configuration maintains a nearly 2:1 sensor-to-mode ratio, yielding a fairly stable $\kappa \approx 20.6$. This represents a two-order-of-magnitude improvement in noise immunity, ensuring that the high-frequency ripples of the Mie manifold can be recovered even under significant experimental stochasticity. As shown in **Figure 10d,** the sensing matrix for this optimized configuration reveals a diffuse energy distribution, ensuring each sensor contributes maximum independent information.

This $\kappa$ can be further improved for more accurate reconstruction performance by utilizing optimized sensors, such as a data-driven basis (SVD). However, using a fixed DCT basis ensures that the sensor array is material-agnostic—the physical sensor positions remain permanent whether the sample is PMMA, PS, or an unknown organic polymer. This physics-neutral approach avoids the overfitting common in machine learning models and provides a robust, universal template for the next generation of sparse, high-performance infrared spectrometers.

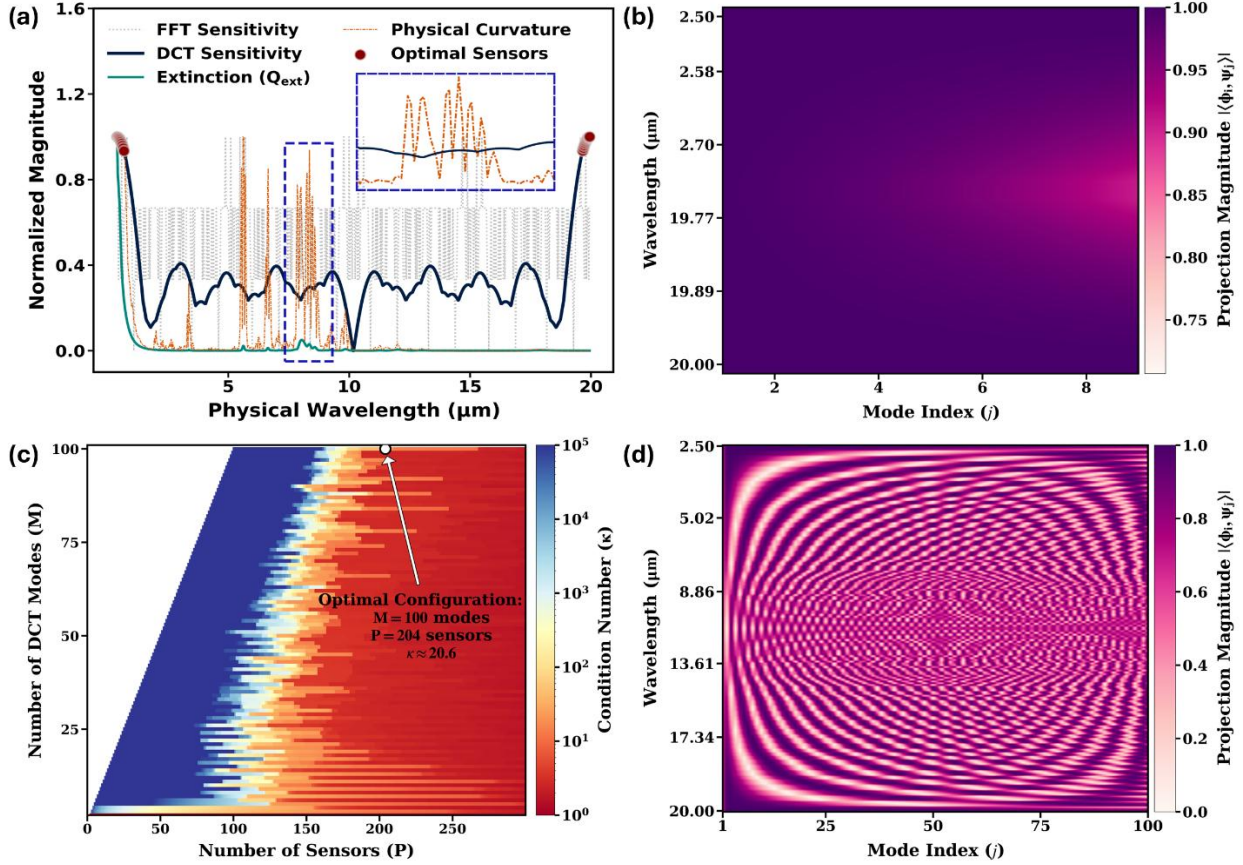

**Figure 10: Optimal Sparse Sensing Configuration and Stability Analysis for Spectral Reconstruction. (a) Multi-Modal Sensitivity and Physical Curvature:** Comparison of DCT (solid navy blue) and FFT (dotted gray) sensitivity scores using $M = 10$ modes and $P = 20$ sensors. The extinction efficiency ($Q_{ext}$) for a particle radius of $0.1\,\mu m$ (solid teal) is plotted alongside its second derivative (dashed orange), representing physical curvature. Sensors (red markers) are optimally placed at the inflection points of the DCT sensitivity profile, which align with regions of high physical curvature, establishing the DCT as a structural proxy for information density. **Inset:** Magnified view of the high-frequency curvature region demonstrating the alignment of optimal sensor locations with resonance features. Since the sensor placement is determined entirely by the mathematical properties of the DCT basis, this strategy provides a universally optimal sampling template that captures structural resonance information without a prior knowledge of material properties or particle size. **(b) Low-Order Sensing Matrix Projection:** Magnitude of the sensing matrix Θ (projection $|\langle \phi_i, \psi_j \rangle|$) for the $P = 20$ sensors and $M = 10$ mode configuration. The heatmap illustrates the wavelength-to-mode mapping, highlighting high projection density at the spectral boundaries used to anchor the global signal envelope. **(c) Stability Phase Transition:** Evaluation of reconstruction stability as a function of DCT mode count ($M$) and sensor budget ($P$). The color scale represents the condition number ($\kappa$), with the optimal operating point identified at $M = 100$ and $P = 204$, achieving a fairly stable $\kappa \approx 20.6$. The sharp color gradient indicates the phase transition between numerically unstable (high $\kappa$) and stable (low $\kappa$) reconstruction regimes. **(d) Optimized Sensing Matrix:** Full projection magnitude for the stabilized configuration ($M = 100, P = 204$). The diffuse interference patterns across the mode-

wavelength space demonstrate the comprehensive, non-redundant information captured across the $2.5 - 20 \ \mu m$ mid-IR range, facilitating high-fidelity spectral reconstruction.

# Conclusion

This work establishes a rigorous information-theoretic framework for quantifying the structural complexity and intrinsic dimensionality of the optical extinction manifold ($Q_{ext}$) in dielectric polymers. By analyzing a diverse library of organic materials, we have demonstrated that the $Q_{ext}$ signal is inherently sparse and governed by a universal, size-dependent complexity profile that transcends specific molecular chemistry. In doing so, we provide a mathematical answer to a 50-year-old question in colloid optics [3] regarding the nature of information-rich regions in extinction; we have quantified the discrete information density of these regions and mapped their universal coordinates across the extinction manifold.

Our findings reveal that for the entire class of dielectric polymers, spectral information density is non-monotonic. We have identified a universal information bottleneck occurring at the onset of the Mie transition regime ($r \approx 0.1 \ \mu m$). At this critical region, spectral entropy and structural disorder reach a global maximum, representing the fundamental physical limit of spectral dimensionality. This bottleneck defines the "worst-case" requirement for any spectroscopic system, necessitating the highest number of degrees of freedom to resolve the interference-driven resonance ripples. While recent theoretical analysis in nonlinear inverse scattering suggests that the number of degrees of freedom should increase with material contrast [10], our results demonstrate that for the dielectric polymer class, the information bottleneck acts as a universal cap on complexity. Notably, our sensitivity analysis confirms that this bottleneck is topologically invariant: even under 10% additive Gaussian noise, the position and structural features of this complexity peak remain fixed. This identifies the bottleneck as a fundamental physical constant of the scattering manifold rather than a transient numerical artifact.

Crucially, this inherent low-dimensionality provides the long-sought explanation for the near-perfect reconstruction of internal properties and geometry in complex polymer mixtures and model biological cells [13, 14, 11, 12]. The ability to resolve 7 individual component spectra from a single mixture profile is not a numerical anomaly; it is a direct consequence of the manifold's structural invertibility. Because the information is efficiently encoded in a sparse set of dominant modes, the system remains observable even under high levels of chemical complexity.

Comparative benchmarking proves that the DCT is the optimal harmonic basis for resolving the Mie manifold, mirroring recent breakthroughs in compressed-domain object detection [30]. While recent experimental benchmarks have established that compressed sensing can accelerate nano-FTIR imaging [5], our results prove that the standard Fourier (FFT) approach is mathematically suboptimal due to inherent spectral leakage. By harmonizing with the non-periodic boundary conditions of extinction data, the DCT achieves a 12-fold reduction in required dimensionality compared to the FFT-based benchmarks reported in the literature, particularly when resolving the fine-grained resonance features of the transition regime. This effectively bypasses the precision-complexity paradox, allowing for high-fidelity reconstruction where standard Fourier techniques become trapped in a regime of diminishing returns and mathematical artifacts.

By bridging the gap between wave scattering and information theory, we transition spectroscopy from a redundant heuristic-based sampling approach to an optimal experimental design. We have demonstrated that 51% (at the Mie transition) to 94% (in the Rayleigh/Geometric regimes) of the data collected by conventional high-resolution spectrometers is redundant. By targeting the information-rich coordinates identified by our framework, we ensure that reconstruction error is limited by the detector's stochastic noise floor rather than basis mismatch.

Ultimately, the DCT framework provides the foundation for a new generation of "thin" spectrometers—computationally lightweight, high-speed instruments capable of resolving the most complex scattering physics with a fraction of the traditional hardware overhead. This paradigm shift holds profound implications for real-time infrared clinical cytology, remote atmospheric sensing, and the development of low-cost, high-precision material characterization systems.

# Author Contributions

P.A. conceived the project, implemented the computational methods, conducted all simulations, performed all the data analysis and authored the manuscript.

# Acknowledgements

I would like to acknowledge my research advisor Prof. Reinhold Blümel for letting me pursue this project independently. I also acknowledge Mr. Arunn Suntharalingam for invaluable research discussions, which contributed significantly to the development of this project.

# Funding

This research received no external funding.

# Conflict of Interest

The author declares no competing interest.

# Data Availability

All data used in this study are drawn from freely available public sources (see corresponding references where applicable). The computational framework relies on the open-source Python packages *PyMieScatt* and *SciPy* (see corresponding references in the manuscript), and all analyses can be reproduced using this software. Detailed step-by-step instructions for reproducing the results are provided in the Methods section of the paper.